# Chloride Ion Erosion of Pre-Stressed Concrete Bridges in Cold Regions

Hongtao Cui [a], Yi Zhuo [b], Dongyuan Ke [c], Zhonglong Li [a], Shunlong Li [a] [*]

a. *School of Transportation Science and Engineering, Harbin Institute of Technology, 73 Huanghe Road, 150090 Harbin, China*

b. *China Railway Design Corporation, 300142 Tianjin, China*

c. *The Seventh Design Institute Co., Ltd, Shanghai Municipal Engineering Design Institute Group, 266001 Qingdao, China*

**Abstract:** The erosion of chloride ions in concrete bridges will accelerate the corrosion of reinforcement, which is an important reason for the decline of bridge durability. The erosion process of chloride ion, especially deicing salt solution in cold regions, is complex and has many influencing factors. It is very important to use accurate and effective methods to analyze the chloride ion erosion process in concrete. In this study, the pre-stressed concrete bridge retired in the cold region was taken as the research object, and the specimens from the whole bridge are obtained by the method of core drilling sampling. The concentration of chloride ion was measured at different depths of the specimens. The process of chloride ion erosion was simulated in two-dimensional space through COMSOL multi-physical field simulation, and compared with the measured results. The simulation method proposed in this paper has good reliability and accuracy.

# 1. Introduction

As one of the important infrastructure for the national socio-economic development,

[*] Correspondence to: Shunlong Li, School of Transportation Science and Engineering, Harbin Institute of Technology, 150090, Harbin, China. E-mail: lishunlong@hit.edu.cn

bridge is the key node of road engineering construction and the main component of urban three-dimensional traffic construction [1-3]. It is also an important reflection of the comprehensive national strength or regional economic development level of the country [4-6]. Compared with other system bridges, pre-stressed concrete bridges have the advantages of short construction period, low construct and maintenance costs [7]. It has become the main bridge form in urban, highway and railway projects. Pre-stressed concrete bridges are mostly of medium and small span, and subjected to repeated action of environment and vehicle load in normal service life [8]. Over time, the material of pre-stressed concrete bridge has deteriorated, and the ultimate durability is insufficient, leading to potential safety hazards.

Compared with other regions, pre-stressed concrete bridges in cold regions are extra eroded by freeze-thaw cycles and deicing salt solutions [9]. According to the investigation, deicing salt can cause serious erosion to concrete bridges in a short period of 10 to 20 years. The erosion of harmful ions in deicing salt is an important reason for the destruction of bridges in cold regions[10][11]. At present, chlorine salt is the main component of deicing salt used in cold regions. With the joint action of freeze-thaw cycle and vehicle load on the bridge, the concrete micro-pore increases[12], the carbonization depth deepens[13], and the reinforcement accelerates corrosion[14]. As a result, the deterioration of various diseases continues to worsen. Therefore, it is particularly important to study the diffusion of chloride ions in pre-stressed concrete in cold regions.

The ways of external harmful ions penetrating into concrete can be divided into many forms, including diffusion, migration, convection, etc. Among them, diffusion is the main mode, which obeys Fick's second diffusion law [15]. Lambert [16] explained the damage mechanism of chloride ion to concrete structure. Once chloride ion penetrates into the surface of steel bar, it will destroy the passivation film on the surface of steel material and accelerate the electrochemical corrosion of steel bar under the combination of water and oxygen. Chiker et al. [17] conducted immersion experiments with sodium sulfate and sodium chloride under the conditions of different types of concrete, water cement ratio and mineral admixtures. The changes of microstructure and mechanical properties of concrete were studied based on scanning electron microscopy.

In terms of the impact of freeze-thaw cycle, Kessler et al.[18] proposed a test method to test the resistance of concrete to freeze-thaw action and chloride ion erosion, and obtained that the service life of concrete structures is accelerated and shortened during the freeze-thaw cycle and chloride ion coupling erosion. Li et al.[19] studied the mesoscopic characteristics of concrete such as pore size, pore distribution, pore structure of interface layer, and concluded that the pore structure of concrete has changed greatly under the action of chloride and freeze-thaw cycles. Jiang [20] et al. studied the corrosion of chloride under the action of freeze-thaw cycles, and also reached the conclusion that the corrosion of concrete is accelerated under the action of freeze-thaw cycles. Suzuki[21], Chen[22] and Tian[23] analyzed the microstructure changes of concrete samples under freeze-thaw damage based on different test methods, and finally reached a similar experimental conclusion that the freeze-thaw cycle will increase the pore size and porosity of concrete.

In terms of the influence of carbonation, Liu et al. [24]proposed that carbonation would lead to changes in the pore structure of concrete materials and thus inhibit chloride ion transmission based on the alternating carbonation and salt spray accelerated corrosion test. Ye et al.[25] pointed out that the carbonation of concrete will cause the chloride ion to re-combine and distribute, making the chloride ion move from the carbonated area to the non-carbonated area. Xiao et al.[26] studied the interaction between carbonation and chloride ion of concrete, and established the interaction model of carbonation and chloride ion erosion of concrete of test type.

In terms of chloride ion erosion simulation, Wang et al.[27] proposed a simplified chloride ion diffusion empirical model considering the time dependence of surface chloride ion concentration based on the results of concrete specimens exposed at different times within 600 days in the tidal zone. Chidiac et al.[28] established the direct relationship between the macroscopic material mechanical properties of concrete and the chloride ion diffusion properties, and proposed a phenomenological model for the analytical expression of chloride ion diffusion with multiple factors. Shafikhani et al.[29] invented a phenomenological multi-scale model to calculate the effective diffusion coefficient of chloride ions in cement mortar and concrete.

Based on the above, the current research generally adopts the accelerated erosion test method in the laboratory. However, due to the huge difference between the laboratory simulation environment and the natural service environment of the bridge, the conclusions obtained by these methods are difficult to be directly applied to concrete bridge structures under real environment. The problems of scarcity of experimental materials, long test period and high cost exist in the physical test of existing bridges, which is currently restricting scholars in various countries to adopt the research method of chloride ion erosion physical test under the real environment.

This paper takes the pre-stressed concrete bridge serviced in the natural cold environment as the research object, and studies the chloride ion erosion mechanism and process from three aspects of theory, experiment and numerical simulation. The meso-structure of concrete was tested and analyzed by X-ray CT scanning technology. Based on the test method of solution titration, the spatial distribution of chloride ion concentration was analyzed, and the key parameters required in the chloride ion erosion simulation of concrete were determined. COMSOL multi-physical field was employed to simulate chloride ion diffusion, and the influence of multiple factors were analyzed on chloride ion erosion.

## 2. Chloride ion erosion theory

### *2.1 Chloride ion transport mechanism*

Generally, chloride ions in concrete mainly include two types: mixed type and penetrating type. Mixed type refers to the concrete composite material wearing chloride ion components, such as mixing water and additives. However, at present, the specification has strictly limited the content of chloride ions in materials, and the harmful ions entering the concrete by this way are relatively small. Another way for chloride ions to enter the bridge concrete structure is that chloride ions in the atmospheric environment, deicing salt and other environments gradually enter the concrete with the increase of the age of the structure. This kind of erosion is the main way of chloride ion corrosion in concrete and cannot be avoided during the service of bridge concrete structures.

As a composite material, concrete will produce capillary pores due to the uneven hydration during the molding process. These pores provide an effective transmission channel for chloride ions in the external service environment of concrete to enter the interior of the structure. For the superstructure of bridges in cold regions, deicing salt solution mainly enters the concrete by diffusion. The diffusion of chloride ions in concrete bridge structures mainly refers to the concentration difference between the outside and the inside of concrete pore solution caused by environmental effects. Therefore, chloride ions will be transported from relatively high concentration to low concentration along the direction of concentration gradient. Because ions as a substance will always conform to the law of conservation of matter, Fick's second law, which has been adopted by most of the chloride ion models which mainly focus on diffusion at present.

## *2.2 Influential factors of chloride ion transport*

The erosion of concrete is affected by many factors. This paper analysed the influence of the concrete's own heterogeneity and the climate characteristics in cold regions on the concrete bridges. The aggregate, interface layer, temperature, freeze-thaw cycle, time effect, carbonation effect and other factors of concrete are summarized and analyzed, and the parameters of different factors were fitted and established.

### 2.2.1 Retardation of coarse aggregate

Coarse aggregate is one of the main components of multiphase composite materials in concrete bridges, which has an important impact on the chloride ion erosion process in concrete. Because coarse aggregate can be regarded as a medium that does not transmit chloride ions, the greater the compactness of aggregate, the greater the retarding effect of chloride ion erosion. At the same time, the larger the volume fraction of aggregate in concrete, the less the content of concrete mortar will be under the same volume, which will play a certain role in inhibiting the diffusion of chloride ions. Based on the aggregate model reconstructed by CT, this paper considers the influence of coarse aggregate on chloride ion transport by directly simulating the volume and shape of aggregate in the finite element method.

### 2.2.2 Influence of interface transition zone (ITZ)

ITZ refers to a thin layer area between the aggregate and cement mortar in the concrete. The porosity of ITZ is higher than that of the adjacent cement mortar. This pore structure, on the one hand, makes the mechanical strength of this region far lower than that of other regions, and is the region where micro-cracks first occur in concrete; On the other hand, ITZ promotes the propagation of harmful substances such as carbon dioxide and chloride ions in concrete. Because the distribution of ITZ is very uneven and it is a decreasing process towards the mortar area, it is difficult to define the boundary between the mortar and ITZ. The thickness of ITZ is about 10 μm~50 μm. The influence on chloride diffusion coefficient is generally about 1.3~16 times.

**2.2.3 Temperature effect**

In terms of temperature effect of chloride ion, the influence model considering the initial temperature and activation energy of concrete is generally adopted. In this study, the influence coefficient expression of temperature on chloride diffusion fitted by Stephen [30] is adopted, and the calculation formula of the influence coefficient is as follows:

$$k_T = (\frac{T}{T_0}) \cdot e^{q(\frac{1}{T_0}-\frac{1}{T})} \tag{1}$$

where $T$ is average absolute temperature of the environment; $T_0$ is reference initial absolute temperature (293 K) and $q$ is the activation constant which related to water-cement ratio and can be calculated according to the following formula:

$$q = 10475 - 10750 \times w/c \tag{2}$$

where $w/c$ is water-cement ratio.

**2.2.4 Carbonization**

Concrete carbonation is one of the common phenomena in various types of concrete failure. Cement hydration products contain many strong alkaline substances such as C-S-H, $Ca(OH)_2$, etc. The environment created by these alkaline substances can produce a dense and stable passivation film on the surface of steel bars, and protect the steel bars from corrosion. However, there are a large number of acidic gas substances dominated by $CO_2$ in the air environment. With the increase of exposure time in the environment, acid gas will continuously enter the pores of concrete and react with alkaline substances in concrete. It will change the alkaline state in the concrete, accelerate the corrosion of steel and reduce the durability of the structure.

In addition, carbonation also has positive and negative effects on the transport of chloride ion: on the one hand, carbonation reaction reduces the content of $Ca(OH)_2$ in concrete, which weakens the binding ability of concrete and chloride ion. The relative proportion of free chloride ion increases, which promotes the transport of chloride ion. On the other hand, due to carbonation, $CaCO_3$ which is difficult to be dissolved in water will be generated, which blocks the transport path of chloride ions in concrete, and thus hinders the transport of chloride ions.

At present, many prediction models of carbonization depth have been proposed, and these models have certain differences in the factors, theory and experimental data. However, at present, most of scholars generally agree that the relationship between the depth of concrete carbonation and the square root of time is proportional [31], and the relationship formula is as follows:

$$x_c = k_c \sqrt{t_c} \tag{3}$$

where $x_c$ is the carbonization depth (mm); $k_c$ is the carbonization coefficient and $t_c$ is carbonization time (year). Cheng et al. [32] tested the chloride diffusion coefficient of concrete at different carbonation depths through rapid carbonation test. In this study, the carbonation influence coefficient of chloride ion diffusion is obtained by polynomial fitting method as shown in **Fig. 1**, and the fitting formula is as follows:

$$k_c = 2.999 \times 10^{-6} x^3 - 1.14 \times 10^{-4} x^2 - 8.723 \times 10^{-3} x + 0.9903 \tag{4}$$

where $x$ represents the depth of specimens.

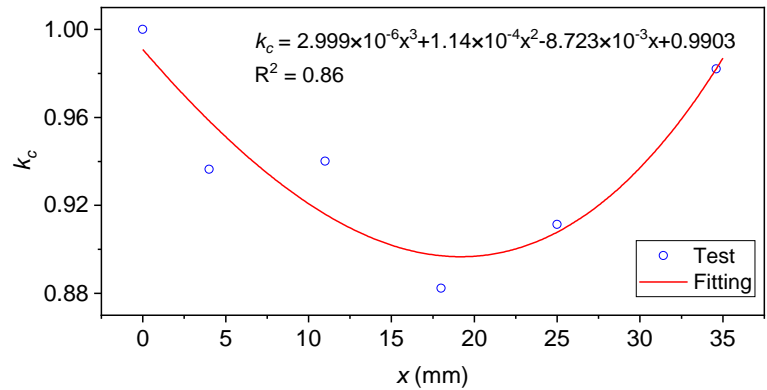

Fig. 1. Carbonization influence coefficient in different depth

**2.2.5 Freeze-thaw cycle**

Freeze-thaw cycle is a unique influence factor in cold regions relative to other regions, and is an important reason for accelerated damage of concrete bridges. The freeze-thaw cycle will significantly increase the internal pores of concrete, and even cause the concrete surface to fall off. These damages caused by freeze-thaw cycles will promote the erosion of chloride ions in concrete bridges. Most of the current research is based on the laws summarized by the rapid freeze-thaw test in the laboratory. These test results cannot be directly applied to the concrete bridge in the real natural environment. Therefore, establishing the conversion relationship between the freeze-thaw cycles in the natural environment is an important prerequisite for the use of laboratory data.

The conversion relationship in this paper is based on the average annual negative temperature days proposed by Wu [33] to calculate the number of freeze-thaw cycles in natural environment, and the calculation results are as follows:

$$n_{act} = \lambda n_f \tag{5}$$

where $n_{act}$ is conversion times of freeze-thaw cycle in natural environment; $n_f$ is annual average negative temperature days and $\lambda$ is correction factor, taken as 0.7. In fact, the freeze-thaw conditions simulated by the laboratory are more severe than the natural environment. The laboratory conditions can be linked to the natural environment through further conversion. The conversion formula is as follows:

$$n_{in} = k_w n_{act} / S \tag{6}$$

where $n_{in}$ is times of freeze-thaw cycle in laboratory; $k_w$ is water content ratio coefficient and $S$ represents damage ratio coefficient between laboratory and natural environment.

Hong [34] fitted the relationship between the number of freeze-thaw cycles and the diffusion coefficient through the test data as follows:

$$D = 0.1064n + 5.4044 \tag{7}$$

where $D$ represents diffusion coefficient of concrete. Combined with the above formula, the conversion relationship between the freeze-thaw influence coefficient in natural environment and the times of freeze-thaw cycles can be established:

$$k_F = 1 + 0.0196876 n_{in} \tag{8}$$

### 2.2.6 Time effect

The diffusion of chloride ion has a strong dependence on the time effect. The main performance is that the concrete will continue to harden with the completion of hydration reaction, and the cement gel will continue to harden, and the internal pores of the concrete will continue to be dense. Thus, the diffusion of chloride ion is hindered. The influence model of time effect is generally established in the form of exponential function[35], and its diffusion

model is as follows:

$$D(t) = D_0 \left(\frac{t_0}{t}\right)^m \tag{9}$$

where $D(t)$ means diffusion coefficient at $t$ moment; $D_0$ is initial diffusion coefficient; $t_0$ is initial reference time, generally 28d and $m$ is time-dependent decay index, taken as 0.264.

### 2.2.7 Concrete binding effect

Not all chloride ions in concrete exist in the form of free state, and concrete materials will have binding effect on chloride ions penetrating into concrete. The adsorbed and combined chloride ions will not have a corrosive effect on steel. The chemical binding effect of concrete is mainly to react with $3CaO \cdot Al_2O_3$ to form Friedel's salt. The chemical reaction is as follows:

$$\begin{aligned}&Ca(OH)_2 + 2NaCl = CaCl_2 + 2Na^+ + 2OH^- \\ &CaCl_2 + 3CaO \cdot Al_2O_3 + 10H_2O \rightarrow 3CaO \cdot Al_2O_3 \cdot CaCl_2 \cdot 10H_2O\end{aligned} \tag{10}$$

In order to express and calculate the effect of chloride ion conveniently, scholars take the content ratio of chloride ion bound state to free state as the binding capacity $R$, and the calculation formula of $R$ can be expressed as:

$$C_f = \frac{1}{1+R} C_t \tag{11}$$

where $C_f$ is free chloride concentration; $C_t$ is total chloride ion concentration and $R$ represents concrete binding capacity, taken between 2 to 4.

### 2.2.8 Construction technique effect

Concrete will have some initial defects due to a variety of uncontrollable factors during the construction and maintenance stage. During the service of the bridge, the concrete material will be subject to shrinkage, edge creep, load and other factors, which will still affect the concrete structure. In order to comprehensively consider these factors into the chloride ion diffusion model, the degradation effect influence coefficient constant is introduced. The value of this coefficient refers to the formula fitted by Wei et al.[36] as follows:

$$k_k = \begin{cases} \left[1000(w/c)^2 - 1050(w/c) + 287\right]/3 & w/c \leq 0.5 \\ 4 & w/c > 0.5 \end{cases} \tag{12}$$

where $k_k$ is deterioration effect influence coefficient.

## 2.3 Theoretical model of chloride ion transport

As a kind of multiphase composite material, concrete itself has certain pore structure. Under the action of the environment, various defects will be produced, such as cracks, voids, pits, etc. Chlorine ions are mainly transported based on these defects and pores. Fick's second law is widely used to describe the diffusion process of chloride ions. The expression of this law is as follows:

$$C_{x,t} = C_0 + (C_s - C_0)\left[1 - erf\left(\frac{x_{cl}}{2\sqrt{Dt}}\right)\right] \tag{13}$$

where $C_0$ is initial concentration of chloride ion; $C_s$ is chloride ion concentration on concrete surface; $erf$ represents error function; $x_{cl}$ means diffusion distance; $D$ is chloride diffusion coefficient and $t$ means diffusion time.

The surface chloride ion concentration refers to the initial value of the concentration of the erosion from the surface of the structure to the interior of the structure. The greater the concentration of chloride ions accumulated on the concrete surface, the stronger the concentration gradient effect and the faster the diffusion rate of chloride ions. In order to fit the surface chloride ion concentration of concrete structures in the real service environment, the modified exponential model established by Kassir [37] test data is used as the expression in this study, and its calculation formula is as follows:

$$C_s(t) = C_{s0} + C_{s\max}(1 - e^{-ct}) \tag{14}$$

where $C_s(t)$ means surface chloride ion concentration at $t$ moment. $C_{s0}$ is initial surface chloride ion concentration; $C_{s\max}$ is final surface chloride ion concentration and $C$ means fitting coefficient.

The relationship between chloride diffusion coefficient and multi-factor influence coefficient can be established by comprehensively considering the above influence factors as follows:

$$D(t) = \frac{1}{1+R} k_F k_T k_k k_c D_0 \left(\frac{t_0}{t}\right)^m \tag{15}$$

where $D_0$ is effective diffusion coefficient of chloride ion at initial time, which can be calculated as: $D_0 = 10^{-12.06 + 2.4(w/c)}$.

# 3. Materials and methodology

## 3.1 Materials

In order to obtain the actual chloride ion erosion and material deterioration damage of concrete bridge structures in cold regions, the core drilling samples from the superstructure of a retired pre-stressed concrete bridge in cold regions were taken as the test objects. The pre-stressed concrete bridge is a simply-supported slab beam, with a total length of 15.96 m and a calculated span of 15.40 m. The thickness of the structural protective layer is 35 mm. The vertical and cross section are shown in **Fig. 2**. The service life of the bridge is 27 years. The concrete grade is C40 and ordinary Portland cement P.O 42.5 is used. The pre-stressed steel tendon is 6 Φ 15.2. The diameter of corrugated pipe is 60 mm, and the embedded depth of pre-stressed steel tendon is 140 mm. The annual average temperature in this region was around 5 °C, and the annual average minimum and maximum temperature were -1 °C and 11 °C respectively. The minimum temperature reached below zero for more than 180 days each year, and the number of days between positive and negative temperatures was about 70 days.

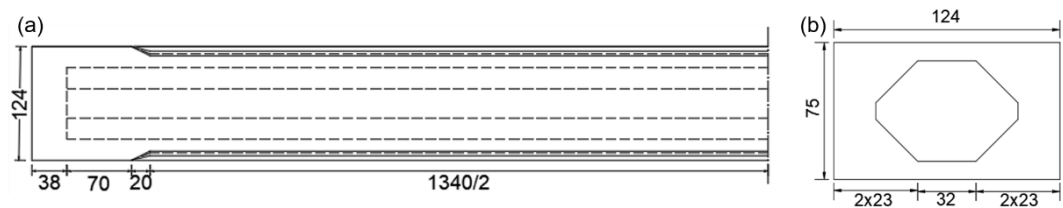

Fig. 2. Diagram of the girder: (a) vertical section; (b) cross section

## 3.2 Methodology

### 3.2.1 Mesoscopic structure analysis

In order to study the internal micro-structure of concrete specimens, the X-ray Computed tomography (CT) made in Germany with the model of Phonix v|tom|x S was used for non-destructive scanning of the specimens. When the X-ray penetrates the test sample, it will be absorbed by the material on the scanning path, resulting in the attenuation of the intensity of the ray. This phenomenon can be explained by the Beer principle. For heterogeneous physics, it is assumed that the object is divided into multiple thin layers. The intensity of X-ray after penetrating the object is as follows:

$$I = I_0 e^{-\int_L \mu_n \Delta x} \quad (16)$$

where $I_0$ represents the initial intensity of X-ray; $u_n$ is attenuation coefficient of thin layer material to X-ray; $L$ is material thickness along the ray direction and $\Delta x$ represents the thickness of thin layer. Working principle of X-ray CT can be seen in **Fig. 3**.

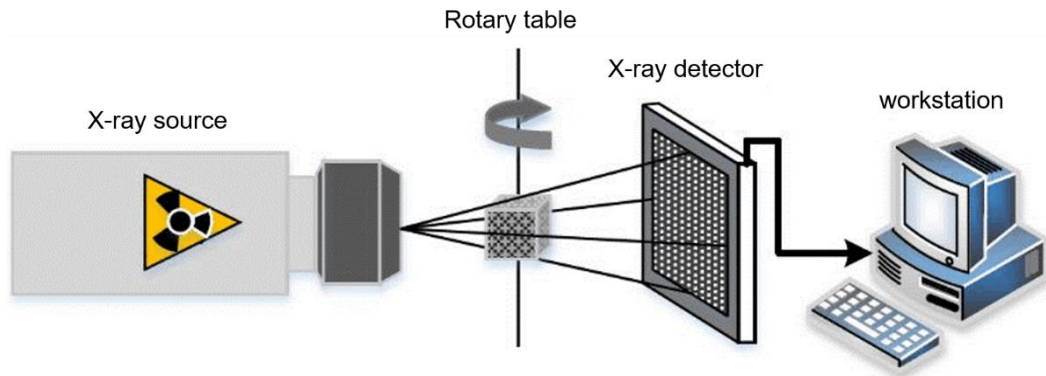

Fig. 3. Working principle of X-ray CT

Six concrete specimens were selected in this paper, and the test voltage and current were 200kV and 110μA, respectively. After the RGB image was converted into a grayscale image, the irrelevant areas such as the external air background of the concrete specimen were removed. The macro-pores and aggregates of CT image were segmented by image threshold segmentation method, and the rest was mortar. The image segmentation process and threshold are shown in **Fig. 4**. According to the 3D reconstruction results of CT after threshold segmentation, the volume account of macro-pores (aperture > 100 μm) was 0.82 %, the average volume ratio of aggregate was 54.03%, and the average volume ratio of mortar was 45.15%. The test results can provide parameters for the subsequent establishment of concrete models.

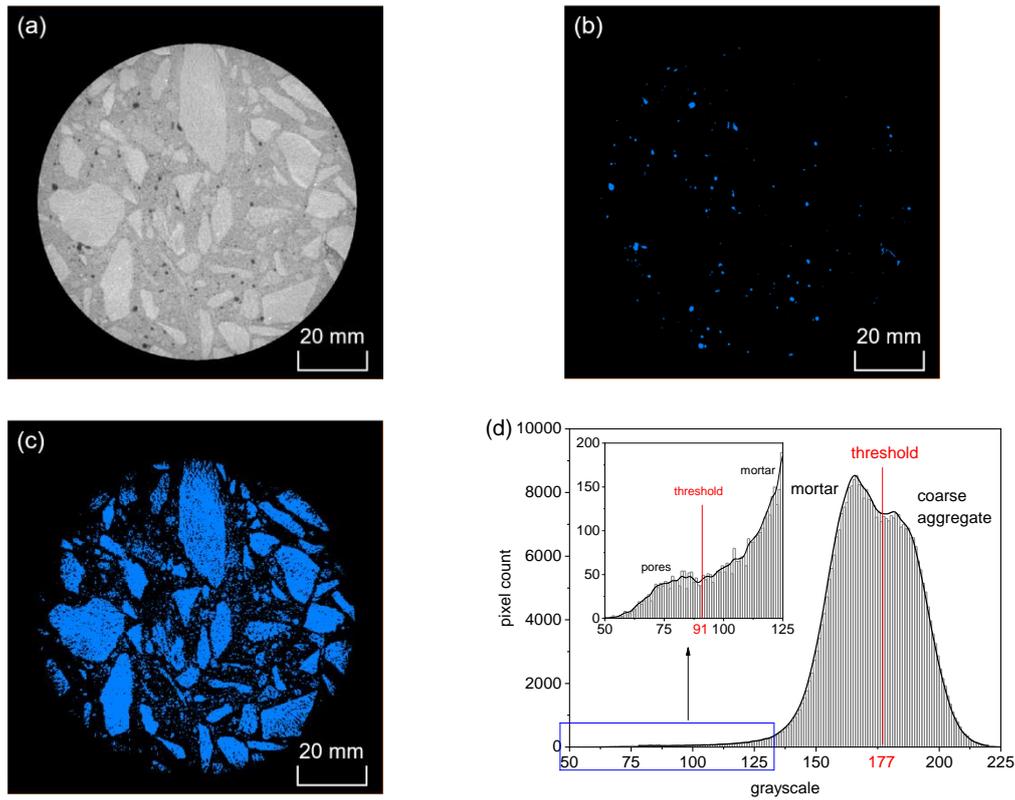

Fig. 4. The image segmentation process: (a) original image; (b) macropores; (c) aggregates; (d) threshold of grayscale

### 3.2.2 Carbonization depth test

In the natural atmospheric environment, pre-stressed concrete bridges will inevitably carbonize with CO2 in the air. In this study, phenolphthalein alcohol reagent was used to spray the sample at the coring site. The carbonation depth of concrete was tested at 10 locations of the girder, and the test results are shown in **Fig. 5**. According to the test results, the average carbonation depth of concrete members is 19.00 mm.

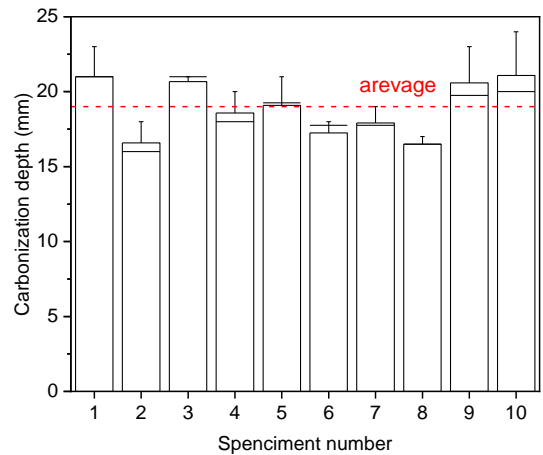

Fig.5. Carbonization depth

### 3.2.3 Chloride ion concentration test

Silver nitrate solution titration method was used to titrate the chloride ion concentration of concrete. The experimental principle is to calculate the content of chloride ion through the reaction between chloride ion and silver ion in the titration solution, and indicate the end point of the reaction through potassium chromate indicator. Titration reaction and indicator reaction are as follows:

$$NaCl + AgNO_3 \rightarrow AgCl\downarrow + NaNO_3 \quad (2)$$

$$2AgNO_3 + K_2CrO_4 \rightarrow Ag_2CrO_4\downarrow + 2KNO_3 \quad (3)$$

The test process of chloride ion concentration distribution in concrete is as follows:

(1) Cut the concrete core specimen into pieces every 1 cm along the depth, and grind the mortar part of each piece with mortar. Grind the mortar powder through 80 μm sieve, and put it into 105 °C oven for drying.

(2) Use a conical flask to soak the dry cement powder in distilled water with a mass ratio of 1:25 for 48 hours. During the immersion process, the conical flask is shaken to fully dissolve the free chloride ion in the powder.

(3) Filter the mixture through qualitative filter paper, and take 20ml of filtrate into a conical flask.

(4) Drop 1-2 drops of phenolphthalein solution standard indicator into the test solution.

Add 0.1mol/L dilute sulfuric acid solution to adjust the extraction solution to neutral. Add 10 drops of standard potassium chromate solution as indicator

(5) Titrate the silver nitrate solution with a titer of 1mg/mL, and shake it while dripping until the color of the extract sample is light orange. Record the milliliter of titrated silver nitrate solution

(6) Calculate the content of free chloride ion in concrete powder according to the volume of titrated silver nitrate solution:

$$C_{cl} = \frac{C_{AgNO_3} \times V_1 \times V_2}{m \times V_3} \times 100\% \tag{4}$$

where $C_{AgNO_3}$ is the titer of silver nitrate solution; $V_1$ is the volume of silver nitrate solution; $V_2$ is the volume of distilled water; $V_3$ is the volume of extracting solution and $m$ represents the quality of mortar powder.

## 4. Chloride ion erosion model of mesoscopic concrete

### 4.1 Establishment of 2D random aggregate background

In this paper, when simulating the diffusion process of chloride ion in concrete, it is considered that chloride ion will not be transmitted inside the aggregate. Fuller ideal grading curve is adopted to add different graded coarse aggregate in proportion to meet the requirements of volume content of coarse aggregate. The content and particle size range of coarse aggregate in real concrete structures are obtained from the aforementioned X-ray CT scanning test results. The aggregate is single graded and the particle size is between 5 mm and 20 mm.

Before generating random polygonal aggregate, randomly generated circular aggregate within the frame of the specimen. Then complete random release. Aggregate coordinates (*x*, *y*) and radius *R* were generated by the "*uniform*" function in Python. Each time a new aggregate was generated, it was judged whether it interferes with the original aggregate. The judgment basis is as follows:

$$\sqrt{(x_n - x_j)^2 + (y_n - y_j)^2} \geq \eta(r_n + r_j) \tag{4}$$

where $(x_n, y_n)$ is coordinate of new aggregates; $(x_j, y_j)$ is coordinate of old aggregates; $r_n$ and $r_j$ are radius of new and old aggregates, respectively. $\eta$ is range influence coefficient, taken as 1.05. The section transition zone can be obtained by extending the circular aggregate outward. The circular aggregate rendering was shown in **Fig. 6 (a)**.

The polygon random aggregate background model was generated based on the circular aggregate. The vertex coordinates of the convex deformation aggregate were determined according to the randomly generated internal angles that meet the angle requirements on the circumference. Then connected each vertex in turn to form the required inscribed convex polygon. As shown in **Fig. 6 (b)**. The interference of the newly generated polygon aggregate was judged according to the interference of the outer circle. If interference occur, it will be regenerated until the generated polygon area met the occupied area of the grading. The polygonal aggregate is generated as shown in **Fig. 6 (c)**.

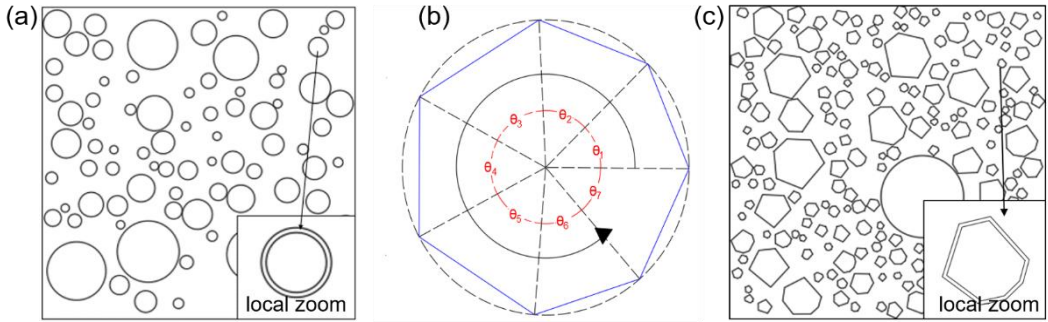

Fig. 6. Aggregate generation process: (a) circular aggregate; (b) polygon generation algorithm; (c) polygonal aggregate

## 4.2 Finite element model

In this paper, the dilute substance transport module in COMSOL Multiphysics was used to simulate the transport process of chloride ion. Import the three-phase aggregate background of concrete generated in Python into COMSOL software. Create the classification selection of concrete aggregate, mortar, interface layer and prestressed steel strand of each material. As shown in **Fig. 7**.

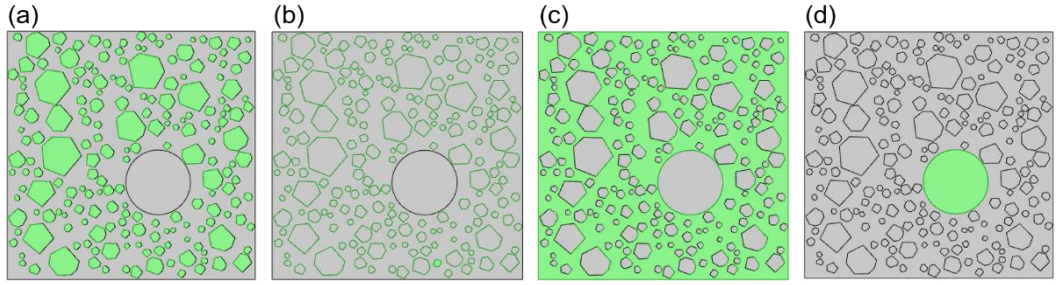

Fig. 7. Division of regions of finite element model: (a) coarse aggregate; (b) ITZ; (c) mortar; (d) pre-stressed tendon

The accuracy and running time of numerical simulation results are closely related to the setting of grid division. Therefore, the calculation time should be reduced as much as possible while the required accuracy is met. Because the concrete aggregate does not participate in the diffusion and transportation of chloride ions, the corresponding domain of the concrete aggregate was removed when dividing the grid. In this study, the free triangular network was used to divide the diffusion area of the remaining chloride ions into grids, as shown in **Fig. 8**.

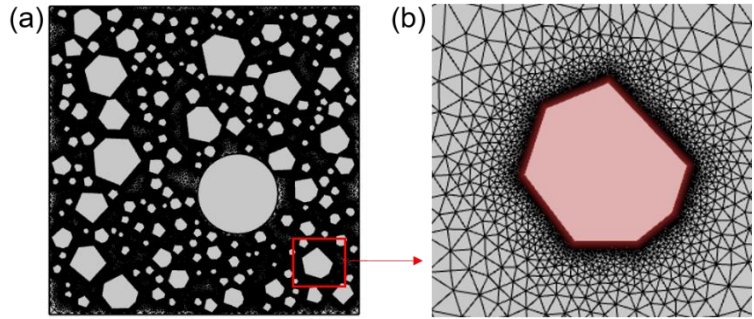

Fig. 8. Model meshing: (a) global meshing; (b) local zoom

According to the theoretical model and test data of chloride diffusion mentioned above, the parameters entered in the finite element model are determined as **Table 1**:

Table 1. Parameters entered in FEM

| Parameter | Value | Parameter | Value |
|---|---|---|---|
| $C_{s0}$ | 0 | $n_{in}$ | 8.1 |
| $C_{smax}$ | 0.37 % | $T$ | 278.3K |
| $C$ | 0.18738 | $q$ | 5175.25 |
| $k_c$ | 3.656 | $k_T$ | 0.3808 |
| $n_{act}$ | 93.38 | $k_k$ | 4.223 |
| $S$ | 11.5 | $R$ | 2.14 |
| $k_w$ | 1 | $D_0$ | $1.30 \times 10^{-11} \text{m}^2/\text{s}$ |

The service life of pre-stressed concrete bridge in this study was 27 years. Therefore, the simulation results of the 27th year of the model were extracted for verification. Compare the theoretical value simulated by the finite element erosion model with the measured value, and the results are shown in the **Fig. 9**. Due to the environmental effects such as scouring on the concrete surface during service, the tested concentration of chloride ion was lower. When chloride ion diffuses in concrete, the theoretical value of chloride ion finite element simulation proposed in this study was in good agreement with the measured value. It is proved that the concrete erosion model and simulation method proposed in this study have good accuracy and reliability.

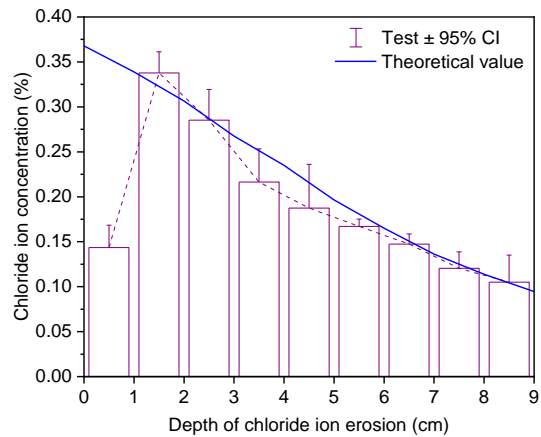

Fig.9. Comparison between test and theoretical values of erosion concentration

The concentration distribution of chloride ion in the erosion time of 5 a, 10 a, 20 a, 30 a, 40 a and 50 a was simulated in COMSOL software, as shown in **Fig.10**. It can be clearly seen that the chloride ion diffuses unevenly locally. The phenomenon of chloride ion gathering, stacking and diffusion path change was obvious due to the direct retardation effect of concrete aggregate and pre-stressed steel strand. These phenomena can also reflect the good simulation effect of the two-dimensional chloride ion erosion finite element model established in this paper.

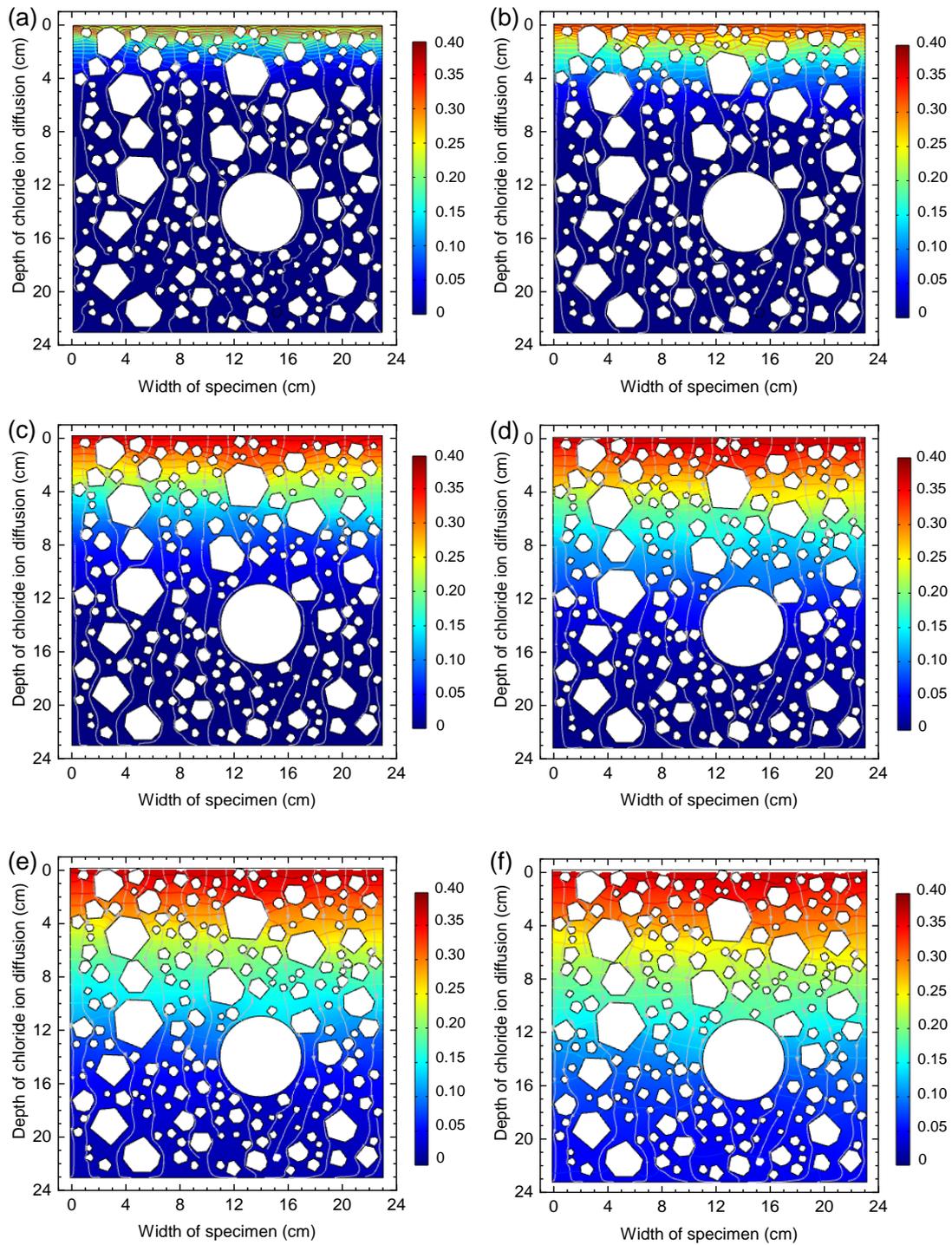

Fig.10. Chloride ion diffusion simulation results in COMSOL: (a) 5a; (b) 10a; (c) 20a; (d) 30a; (e) 40a; (f) 50a

Take the maximum value of two-dimensional section at different depths as the most adverse effect of chloride ion due to the non-uniformity of concrete. The calculation results are shown in **Fig. 11**. It can be seen that the maximum chloride ion at different depths caused by the heterogeneity of concrete was not uniformly and smoothly decreased. In this paper, 0.06% is taken as the critical concentration of chloride ion for steel strand corrosion. From the chloride ion erosion curve in the 27th year, the chloride ion concentration on the surface of pre-stressed tendon has reached the critical concentration of chloride ion during the service life. If factors such as concrete surface peeling and cracks are considered, the actual corrosion of pre-stressed steel strand will be more serious. This is consistent with the phenomenon that the bearing capacity of the pre-stressed concrete bridge is insufficient and it is retired in advance. From the predicted results of chloride ion concentration in the 50th year, it can be seen that the chloride ion concentration at the deepest position of the steel strand also exceeds the critical chloride ion concentration. The concentration at the shallowest position of the steel strand reaches more than 3 times of the critical concentration. Therefore, the impermeability of the bridge concrete cannot meet the expected design life.

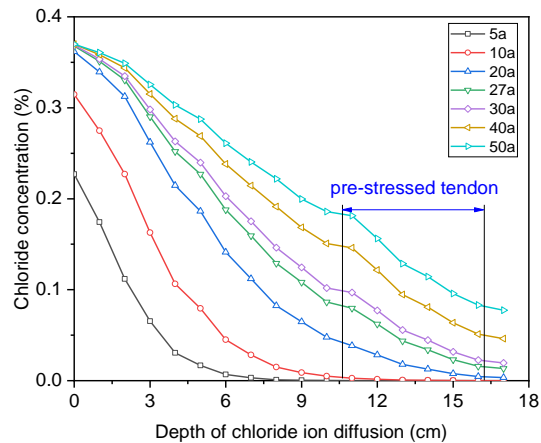

Fig.11. Chloride ion diffusion concentration at different times

## 5. Conclusion

In this paper, the chloride ion erosion mechanism and related influencing factors of pre-stressed concrete bridges in cold regions were analyzed, and a multi-parameter chloride ion erosion theoretical model of chloride ion changing with time in concrete was established. The deterioration of factors related to chloride ion corrosion of pre-stressed concrete bridges in real service environment was tested. Under the condition of considering the change of various influencing factors, the chloride ion erosion condition of different years was simulated and predicted using COMSOL software. The conclusions are as follows:

(1) It was tested that the volume ratio of macro-pores (aperture > 100 μm) was 0.82 %, the average volume ratio of aggregate was 54.03%, and the average volume ratio of mortar was 45.15 %. The carbonation depth of the concrete cover of the bridge is about 1.90 cm in the servicing time of 27 years.

(2) The distribution of free chloride ion concentration was tested by silver nitrate solution titration method, and the depth change rule of free chloride ion was obtained. From the tested concentration change curve, it can be seen that the impact of the environmental effect of chloride ion in the cold regions had a relatively serious erosion effect on the bridge.

(3) Combined with the measured concrete bridge data, the chloride ion concentration erosion distribution of the bridge was simulated by COMSOL software. The correctness of the erosion model was verified by comparison, and the conclusion that the chloride ion concentration on the surface of the steel strand has reached its critical chloride ion concentration was obtained. Based on this model, the distribution of chloride ion erosion in 50 years was predicted. In the 50th year, the maximum concentration on the surface of steel strand has reached more than 3 times of the critical concentration. The bridge beam cannot meet the design service life requirements.

## List of abbreviations

ITZ: Interface transition zone; CT: Computed tomography

## Acknowledgements

The authors appreciate the National Natural Science Foundation of China and China Railway Design Corporation R&D Program for support of this research.

## Declarations

**Ethical Approval**

Yes.

**Consent for publication**

Yes.

**Competing interests**

The authors declare that they have no competing interests.

**Authors' contributions**

**Hongtao Cui:** methodology, investigation, writing – original draft and visualization. **Yi Zhuo:** resources. **Dongyuan Ke:** investigation, software, data curation. **Zhonglong Li:** resources and supervision. **Shunlong Li:** conceptualization, writing – review and editing, supervision, project administration and funding acquisition.

**Funding**

Financial support for this study was provided by NSFC [52278299] and China Railway Design Corporation R&D Program [2020YY340619, 2020YY240604]

**Availability of data and materials**

The data and code are available upon request.

# References

[1] Huang Y, Beck JL, Li H (2017) Bayesian system identification based on hierarchical sparse Bayesian learning and Gibbs sampling with application to structural damage assessment. Comput Method Appl M 318:382-411

[2] Li SL, Wei SY, Bao YQ, Li H (2018) Condition assessment of cables by pattern recognition of vehicle-induced cable tension ratio. Eng Struct 155:1-15

[3] Li H, Lan CM, Ju Y, Li DS (2012) Experimental and Numerical Study of the Fatigue Properties of Corroded Parallel Wire Cables. J Bridge Eng 17:211-220

[4] Li H, Mao CX, Ou JP (2008) Experimental and theoretical study on two types of shape memory alloy devices. Earthq Eng Struct D 37:407-426

[5] Bao YQ, Shi ZQ, Beck JL, Li H, Hou TY (2017) Identification of time-varying cable tension forces based on adaptive sparse time-frequency analysis of cable vibrations. Struct Control Hlth 24: e1889

[6] Li SL, Zhu SY, Xu YL, Chen ZW, Li H (2012) Long-term condition assessment of suspenders under traffic loads based on structural monitoring system: Application to the Tsing Ma Bridge. Struct Control Hlth 19:82-101

[7] Kim J, Park S (2020) Field applicability of a machine learning-based tensile force estimation for pre-stressed concrete bridges using an embedded elasto-magnetic sensor. Struct Health Monit 19:281-292

[8] Gao C, Zong ZH, Lou F, Yuan SJ, Lin J (2022) Load model experiment of prestressed concrete continuous girder bridge subjected to explosion above the deck. China J. Highw. Transp 35: 106-114 (in Chinese)

[9] Wei Y, Guo W, Wu Z, Gao X (2020) Computed permeability for cement paste subject to freeze-thaw cycles at early ages. Constr Build Mater 244:118298

[10] Wang B, Wang F, Wang Q (2018) Damage constitutive models of concrete under the coupling action of freeze–thaw cycles and load based on Lemaitre assumption. Constr Build Mater 173:332-341

[11] Bharadwaj K, Glosser D, Moradllo MK, Isgor OB, Weiss WJ (2019) Toward the prediction of pore volumes and freeze-thaw performance of concrete using thermodynamic modelling. Cement Concrete Res 124:105820

[12] Tian Z, Zhu X, Chen X, Ning Y, Zhang W (2022) Microstructure and damage evolution of hydraulic concrete exposed to freeze–thaw cycles. Constr Build Mater 346:128466

[13] Hussain S, Bhunia D, Singh SB (2017) Comparative study of accelerated carbonation of plain cement and fly-ash concrete. J Build Eng 10:26-31

[14] Otieno M, Beushausen H, Alexander M (2016) Chloride-induced corrosion of steel in cracked concrete – Part I: Experimental studies under accelerated and natural marine environments. Cement Concrete Res 79:373-385

[15] Collepardi M, Marcialis A, Turriziani R (1972) Penetration of Chloride Ions into Cement Pastes and Concretes. J Am Ceram Soc 55:534-535

[16] Lambert P, Page CL, Vassie PRW (1991) Investigations of reinforcement corrosion. 2. Electrochemical monitoring of steel in chloride-contaminated concrete. Materials & Structures 24:351-358

[17] Chiker T, Aggoun S, Houari H, Siddique R (2016) Sodium sulfate and alternative combined sulfate/chloride action on ordinary and self-consolidating PLC-based concretes. Construction & Building Materials 106:342-348

[18] Kessler S, Thiel C, Grosse CU, Gehlen C (2017) Effect of freeze-thaw damage on chloride ingress into concrete. Mater Struct 50: 121

[19] Li B, Mao J, Nawa T, Liu Z (2016) Mesoscopic chloride ion diffusion model of marine concrete subjected to freeze-thaw cycles. Constr Build Mater 125:337-351

[20] Jiang WQ, Shen XH, Xia J, Mao LX, Yang J, Liu QF (2018) A numerical study on chloride diffusion in freeze-thaw affected concrete. Constr Build Mater 179:553-565

[21] Suzuki T, Ogata H, Takada R, Aoki M, Ohtsu M (2010) Use of acoustic emission and X-ray computed tomography for damage evaluation of freeze-thawed concrete. Constr Build Mater 24:2347-2352

[22] Chen J, Deng X, Luo Y, He L, Liu Q, Qiao X (2015) Investigation of microstructural damage in shotcrete under a freeze-thaw environment. Constr Build Mater 83:275-282

[23] Tian W, Cheng X, Liu Q, Yu C, Gao F, Chi Y (2021) Meso-structure segmentation of concrete CT image based on mask and regional convolution neural network. Mater Design 208:109919

[24] Liu J, Qiu Q, Chen X, Xing F, Han N, He Y, Ma Y (2017) Understanding the interacted mechanism between carbonation and chloride aerosol attack in ordinary Portland cement concrete. Cement Concrete Res 95:217-225

[25] Ye H, Jin X, Fu C, Jin N, Xu Y, Huang T (2016) Chloride penetration in concrete exposed to cyclic drying-wetting and carbonation. Constr Build Mater 112:457-463

[26] Xiao QH, Li Q, Guan X, Zou YX (2018) Prediction model for carbonation depth of concrete subjected to freezing-thawing cycles. IOP conference series. Materials Science and Engineering 322:22048

[27] Wang Y, Wu L, Wang Y, Li Q, Xiao Z (2018) Prediction model of long-term chloride diffusion into plain concrete considering the effect of the heterogeneity of materials exposed to marine tidal zone. Constr Build Mater 159:297-315

[28] Chidiac SE, Shafikhani M (2019) Phenomenological model for quantifying concrete chloride diffusion coefficient. Constr Build Mater 224:773-784

[29] Shafikhani M, Chidiac SE (2020) A holistic model for cement paste and concrete chloride diffusion coefficient. Cement Concrete Res 133: 106049

[30] Amey SL, Johnson DA, Miltenberger MA, Farzam H (1998) Predicting the service life of concrete marine structures: An environmental methodology. Aci Struct J 95:205-214

[31] Yu ZW, Xiao MT, Liu P, He PF, Fan J (2017) Overview and prospect on prediction models for carbonization depth of concrete. Paper presented at the 26th National Academic Conference on Structural Engineering 33: e0074 (in Chinese)

[32] Cheng Y, Sun JY, Zheng M (2017) Experimental study on chloride diffusion under the effect of flexural tensile load and carbonization coupling. Bulletin of The Chinese Ceramic Society 36:1700-1705 (in Chinese)

[33] Wu HR, Jin WL, Yan YD, Xia P (2012) Environmental zonation and life prediction of concrete in frost environments. Journal of Zhejiang University (Engineering Science) 46: 650-657 (in Chinese)

[34] Hong L, Tang XD (2011) Influence of freezing-thawing cycles and curing age on chloride permeability of concrete. Journal of Building Materials 14:254-256 (in Chinese)

[35] Pack SW, Jung MS, Song HW, Kim SH, Ann KY (2010) Prediction of time dependent chloride transport in concrete structures exposed to a marine environment. Cement Concrete Res 40:302-312

[36] Wei J, Gui ZH, Wang YL (2005) Modeling on predicting steel corrosion rate in concrete. Journal of Wuhan University of Technology 27: 45-47 (in Chinese)

[37] Kassir MK, Ghosn M (2002) Chloride-induced corrosion of reinforced concrete bridge decks. Cement Concrete Res 32:139-143